\font\sf=cmss10 at 10pt 

\font\sssf=cmss10 at 6pt

\font\bb=msbm10 at 10pt
\font\bbs=msbm10 at 8pt

\font\rmss=cmr10 at 7pt

\font\fraks=eufm10 at 8 pt
\font\cal=cmsy10 at 10pt
\font\cals=cmsy10 at 8pt

\def\0#1{\mbox{\rm #1}}
\def\1#1{\mbox{\bb #1}}
\def\2#1{\mbox{\bf #1}}
\def\3#1{{\mathcal #1}}
\def\4#1{\mbox{\cals#1}}
\def\5#1{\mbox{\sf #1}} 
\def\6#1{\mbox{\sssf #1}}
\def\7#1{\mbox{\fraks #1}}
\def\8#1{\mbox{\rmss #1}}
\def\9#1{\mbox{\bbs #1}}

\def\BEn{\begin{enumerate}}

\def\EEn{\end{enumerate}}

\def\BE{\begin{equation}}

\def\EE{\end{equation}}

\def\BEA{\begin{eqnarray}}

\def\EEA{\end{eqnarray}}

\def\io{\iota}

\def\Si{\Sigma}

\def\sq{{\rule[0pt]{8.0pt}{.5pt}\kern-8.5pt
\rule[0pt]{.5pt}{8.5pt}
\rule[8.1pt]{8pt}{.5pt}
\rule[0pt]{.5pt}{8.5pt}}}

\def\sqs{{\rule[0pt]{4.2pt}{.5pt}\kern-4.7pt 
\rule[0pt]{.5pt}{4.7pt}
\rule[4.3pt]{4.2pt}{.5pt}
\rule[0pt]{.5pt}{4.7pt}}
}

\def\apo{\mbox{\rm '}}

\def\dag{\dagger}

\def\adj{{^{\dag}}}

\def\lfl{{\lfloor\kern-5pt\lfloor}}
\def\rfl{{\rfloor\kern-5pt\rfloor}}

\def\x{\times}

\def\ox{\otimes}
\def\OX{\bigotimes}

\def\bx{\kern2pt\sqs\kern-8.55pt\x}

\def\from{\kern-2pt\leftarrow\kern-2pt}

\def\pd{\partial}

\def\Bar{\kern5pt{\rule[-2.5pt]{.6pt}{9.5pt}}\kern5pt}
\def\bra{\langle}

\def\ket{\rangle}

\def\Cliff{\mathop{{\rm Cliff}}\nolimits} 

\def\Endo{\mathop{\rm Endo}\nolimits} 

\def\Re{\mathop{\hbox{\rm Re}}\nolimits}

\def\SO{\mathop{\mbox{\rm SO}}\nolimits}

\documentclass[11pt]{article}
\usepackage{graphicx}
\usepackage{amssymb}
\usepackage{makeidx}
\usepackage{graphicx}
\begin{document}
\title{
\huge{Elementary operations}\footnote {Based on a talk given at the
\emph{5th International Quantum Structure Association Conference,
Cesena, Italy, 2001.}To be published in the International Journal of
Theoretical Physics.}}
\author{\bf James Baugh,
Andrei Galiautdinov, David Ritz Finkelstein\footnote{e-mail:
\texttt {df4@mail.gatech.edu}},
\\and \bf Mohsen Shiri-Garakani\\ {\normalsize School of Physics,
Georgia Institute of Technology}\\
{\normalsize and}\\
\bf Heinrich Saller\\
{\normalsize Heisenberg Institute, Max Planck Institute of Theoretical Physics, Munich}
}
\date{}
\maketitle

\parindent=0pt

\abstract{
A Clifford algebra over the binary field $2=\{0,1\}$ is a second-order
classical logic that is substantially richer than Boolean algebra.
We use it as a bridge to a
Clifford algebraic quantum logic
that is richer than the usual Hilbert space quantum logic
and admits iteration.
This leads to
a higher-order Clifford-algebraic logic.
We formulate a toy Dirac equation with this logic.
It is exactly Lorentz-invariant,
yet
it approximates the usual Dirac equation as closely as desired
and all its variables have finite spectra.
It is worth considering as a Lorentz-invariant improvement on lattice space-times.
}
\section{Quantum hierarchy}

The Hilbert-space lattice logic is obviously insufficient
for quantum physics.
It lacks hierarchy.
It is quantum on one level,
the  next level,
on which we talk about predicates,
not quanta,
 is classical and distributive.
We cannot describe a spin  completely,
but we assume that we can describe
a predicate of the spin completely
because the polarizers that control it are practically classical.
The lattice logic is not yet a quantum {\em logic}
but only a quantum first-order predicate algebra,
with classical higher levels.

When we use elementary particles
as probes, say of space-time structure,
this one-tier-quantum logic is unsatisfactory.
There are  experimental
indications of something like the set-theoretic hierarchy
in physics.
In the first place, space-time --- which must ultimately be quantum ---
appears
at least 4-dimensional
to the physicist.
This implies at least four hierarchic levels:
point $\in$ line $\in$ plane $\in$ volume $\in$ space-time.
Then particles and fields are are described by functions on time
or space-time.
This puts them still higher in the hierarchy; and
by at least two levels,
if a function is indeed a  set of pairs.
Finally  quantum fields are built over single-quantum
theories.  This adds at least  another level.
Finally, Lagrangians are functions of quantum fields,
adding at least another level.
A simple physics (we assume) is quantum at every level,
though the experimenter may
may be described classically under low resolution.
Therefore we seek a quantum correspondent to set theory.

Von Neumann understood this well.
His Ph. D. thesis
was already on a hierarchic logic.
Von Neumann  explicitly  posed the problem
of a quantum set theory
that we may have solved here.

\section{Quantum cosmos}

The standard quantum theory represents the operations on a given quantum system
\5S
that can be carried out by an experimenter \5T
by operators on a Hilbert space $H(\5S)$
associated with \5S up to isomorphism.
To unify physics along this line one wants a space
whose operators represent
ideally possible  quantum operations
by any experimenter on on any system.
We  see no alternative but this
commonly made ``\5U assumption:"
\BE\mbox{All  systems and experimenters
are  subsystems of one  cosmic system \5U}.
\EE

We cannot determine the cosmos \5U sharply.
Strictly speaking its existence is metaphorical, not operational.
We must imagine many possible approximate factorizations
$\5U\approx \5S \ox \5T(\5S)$ into system and exosystem,
the exosystem \5T(\5S) comprising
everything not part of \5S, including an experimenter.

Sometimes the factorization into system and experimenter
is called the Heisenberg
cut,
as though Heisenberg had discovered the distinction between
map and territory.
This is naive.
We cannot  trace the map-territory distinction back to its remote
origins but
it was clearly drawn by Plato, Boole, and Peirce
well before quantum theory.
It is only that
physicists did not do much measurement theory
during the golden age of naive classical physics.
Instead they assumed, usually implicitly,  that
the system  and its mathematical
model  were isomorphic, especially in their logics.
In quantum theory
logic is no longer scale-free.
Microcosmic spins (for example) have a different
logic than macroscopic mathematical systems.
It is no longer safe to identify map with territory, even
for calculations.

The \5U assumption renounces operationality.
There is no actual
experimenter to observe \5U sharply
and give operational meaning to its states.
We take much the viewpoint
in quantum physics
that Laplace explicitly took in classical:
that of a purely metaphorical Cosmic Experimenter  or CE
controlling the entire cosmos
with maximal sharpness.
The CE sees our little measurements on the system as reversible interactions
between us and the system.
Each experimenter
partitions the cosmos \5U into system \5S and metasystem \5T(\5S).
We imagine that the metaphorical
CE sees all these as subsystems of one quantum system \5U, the cosmos.
Our CE differs from Laplace's supreme intelligence, however,
in that
our CE is a quantum relativistic experimenter,
while Laplace's is
a classical experimenter.
Instead of Laplace's cosmic phase space
our CE  has a cosmic Hilbert space $H(\5U)$
with cosmic initial states $|\alpha\ket\in H(\5U)$
and dual final states $\bra \omega|\in \dag H(\5U)$,
the dual Hilbert space.
The cosmic Hilbert space  factors into the operational Hilbert space $H(\5S)$
of the usual theory and a mostly inaccessible Hilbert space $H(\5T(\5S))$
for the exosystem.

We must not imagine that the CE can know \5U ``as it is.''
We may imagine that the CE  inputs the cosmos with a process $|\alpha\ket$
of her choice
before we begin our experiments
and outtakes it
with a bra $|bra \omega|$ of her choice
after we finish.
She works with probability amplitudes $\bra\omega|U|\alpha\ket$ for \5U
transitions
much as we do for atomic  transition.
Thus CE  measurements cover and replace ours.
Von Neumann already showed
that such a higher-level experimenter
can make observations consistent with ours.

One  can then describe thermodynamically irreversible measurements by \5T on \5S
as partly known invertible dynamical transformations of the  meta-system
$ \5U=\5S\ox \5T$.
The \5U assumption puts all actual experimenters,
with all possible partitions,
into one algebraic theory,
extending relativity
beyond what is possible for
a truly operational theory.
The  quantum cosmos
is not a new concept for quantum physics.
Bohr
at first rejected angrily
the idea of a quantum universe,
and later advocated such a
higher relativity \cite{BOHR}.
Quantum  field theorists already
take the viewpoint of a CE implicitly.
The person and apparatus of any actual physicist
are just  condensations in  the same
fields that the physicist is treating,
and yet are lacking in the vacuum state of quantum field  theory,
which is Poincar\'e-invariant.
That state must be the vacuum state of \5U for
the CE.
In introducing the CE
we simply make explicit
what  every field theorist
has implicitly or explicitly  done since Dirac's
quantum electrodynamics.

We distinguish therefore between
operational theories,
which have  unitary groups limited to
transforming the system but not the experimenter
and the cosmos,
and operation theories,
whose operations may radically change the entire cosmos
but
are not operational
because they cannot be carried out by any experimenter
in the cosmos.
The same abstract algebraic structure can be interpreted either way.
The metaphorical operations of the
CE on the universe, including dynamical developments
of \5U,  we call
\5U operations.
Those
that we can actually carry out on the system S,
described from our viewpoint,
we  call \5S operations.

\subsection{Group regularization}

Segal pointed out that non-semisimple groups are unstable in an important sense
and  proposed to replace them with simple groups
by making small changes in their commutation relations.
A Lie group $G$ with Lie algebra Lg $G$  is (Segal-) unstable if any neighborhood of its Lie product
\BE
\times: \mbox{Lg }G\ox \mbox{Lg }G\to \mbox{Lg }G
\EE
(in the topology of the manifold of tensors of type $\mbox{Lg }G\ox \mbox{Lg }G\to \mbox{Lg }G$ that obey the Lie product conditions)
contains non-isomorphic products.
That is, a group is unstable if the smallest change in its commutation relations suffices to
change it to another group.
And such a change also serves to stabilize it against further changes of that kind.

Indeed, many major discoveries of physics of the last century
have had just this form.

The group structure or algebraic structure of a quantum theory determines the quantum theory as follows.
Given the unitary group $U=U(H)$ of a Hilbert space $H$, the Hilbert space $H$ is uniquely determined up to isomorphism by the condition that its isometry group be $U$.

We call this regularization process group {\em flexing}\/,
since it introduces a curvature into the group manifold.
The inverse process
that returns to the singular unstable theory is {\em group flattening}\/.
Group contraction and its inverse, group expansion, are special cases
\cite{GILMORE}\/.
The simple Lorentz group needs no regularization, but the Poincar\'e group
and the Heisenberg group are still singular.
Their most economical regularizations are orthogonal groups, whose
representations are efficiently constructed using Clifford algebra.

If ultimately by such small changes in the commutation relations
we arrive at a simple  Lie group for the
quantum theory, it will be a finite quantum theory,
with a discrete bounded spectrum for every observable.
Its Hilbert space will be finite dimensional.

\subsection{Clifford algebra concepts}

We use the following
Clifford algebra concepts and notation in what follows.

A  Clifford ring $C$
consists of all polynomials in elements of  a linear space $C_1$ called vectors,
with a commutative unital ring  of coefficients $C_0$  called scalars,
obeying
the {\em Clifford law}\/:
\BE
\mbox{The square of any vector is a scalar.}
\label{CLIFFORDLAW}\
\EE

For any vector $v$ of $C$ we define
\BE
\|v\| =: v\adj\ v =:v^2:= \dag_{mn}v^mv^n.
\EE
This  makes $C_1$ a quadratic space over the
coefficient ring $C_0$.
If $C_0$ is a field we call the Clifford ring a Clifford algebra.

A free Clifford algebra is  defined
by its quadratic space $V=C_1$ over the coefficient field $C_0$
and is written $C=\22^V$.
If  $C$
is isomorphic to the endomorphism
algebra of a module $S$,  over some ring possibly different from $C_0$,
$C(V)\sim S\ox \dag S$,
then one calls $S)$ a
 {\em spinor} module for $\Cliff V$,
and we write this improperly as
$S=\Si C$.

The spinor space
 $S$ supports a projective representation of $\SO(V)$.
This  representation of $\SO(V)$
is reducible.
The proper
irreducible  subspaces
of $S$ are also called  spinor spaces .

We write $\5T: x\mapsto x^{\6T}, \; xy\mapsto y^{\6T}x^{\6T}$
 for {\em transposition}\/,
the natural anti-automorphism
$C\to C$ fixing every
first-grade Clifford element $  e=e^{\6T}$.
We write $\5C: x\mapsto x^{\6C}$
for Clifford {\em conjugation},
the automorphism of $\Cliff V$ that changes the sign
of first grade elements:
$e^{\6C}=-e$.
Then $\5H:=\5T\5C$ is a natural anti-automorphism
that changes the signs
of the first-grade Clifford elements: $e^{\6H}= -e$\/.
The {\em four-group} $G_4(C)$ of the Clifford algebra
consists of the mappings $\5I, \5T, \5C, \5H: \Cliff V\to \Cliff V$\/,
\5I being the identity mapping.

Four natural quadratic forms on the Clifford algebra $C$
 are the $\|z\|_{\6x}:=\Re (x^{\6x}x)$
where $\5x\in G_4(C)$ is any of the four-group elements.
We use the one invariant under the largest group
of inner automorphisms of $C$,    $\|z\|:=\|z\|_{\6I}$\/.

The  Clifford group $\0C\0G\; C$of $C$
is
the group of
invertible elements of  $C$.

For any algebra $A$ and any element $z\in A$,
$\5L z$ and $\5R z$ designate
 the linear operators  $A\to A$ of
left-multiplication and right-multiplication
by $z$:
$(\5L z)x=zx, \quad (\5Rz)x = xz$.
The associative law just states  that $\5Lx$ and $\5R y$
commute for arbitrary $x$ and $y$.

\section{Clifford logic}

\subsection{Clifford classical logic}

In classical physics one models classes or predicates of the system by subsets
of its phase space.
We shall model a quantum class or predicate
by an aggregate, now
a quantum aggregate of quantum systems,
not classical.
We use this logic to describe physical processes within \5U, and these
are assumed to be reversible (if only by the CE).
Therefore we seek a reversible logic $L(\5S)$ for the physical system \5S.
Its basic logical operation is to be a group operation.
There are only two classical logical operations
 that are group operations:
$\equiv$
(material equivalence, ``neither or both'')
and its negation ${\not\equiv}$, usually called XOR
(``either and not both''), designated here by $\sqcup$.
Their group identities are FALSE and TRUE respectively.
For the sake of familiarity we fix the logic group operation
as $\sqcup$.
 Then we must represent FALSE by the multiplicative identity
$1\in L(\5S)$.

This corresponds to representing
the vacuum by 1 in Fock's theory of fermions or bosons.
Do not confuse this $1\in L(\5S)$ representing FALSE,
 with the linear operator
$I\in A(\5S)$
representing TRUE or GO in the lattice logic.

Every classical predicate can be expressed in classical (finite)
logic as a XOR of disjoint atoms.
We therefore represent the quantum atoms first.

We generate the classical XOR group with the relations
\BE
\label{eq:XOR}
\gamma\; \mbox{\rm  XOR }\gamma = \gamma^2=1, \quad \gamma\gamma'=\gamma'\gamma
\EE
 for different atom $\gamma\ne \gamma'$ of the predicate lattice.
The XOR group is graded
by
the number of atoms in a predicate.

 Over the binary field $2=\{0, 1\}$,  anticommuting is the same as commuting.
 The group (convolution) algebra of the XOR group is evidently a Clifford algebra
$\Cliff(\5S, 2)=: \22^S$
over 2 .
We use it as
a guide to the Clifford quantum  logic.
We call its cliffors {\em binors} for brevity.

One can form the binor algebra $\Cliff(\5S, 2)$
of any finite set $\2S$ of distinct objects $s_0, s_1, \dots, s_N$.
We form  unit sets $\gamma_n:=\iota s_n$ of grade 1 and
provide them
 with the binary sum operation
 $+:\iota \apo \2S\x \iota \apo \2S\to \iota \apo \2S$ and
 product $\sqcup: \iota \apo \2S\x \iota \apo \2S \to \iota \apo \2S$.
 To keep the levels straight, it is important that the binor
representing a set is the product of
its unit subsets,
not of its elements.

 The general basis element of $\22^{\2S}$ is a monomial in the variables
 $\gamma_n\in \iota \apo \2S$.
  A variable whose square is 1 we call {\em uniquadratic}\/.
Then the $\gamma_n$  make up
a first-grade basis of $N$ (anti)commuting
uniquadratic generators  of $\22^S$
unique up to order.

\subsubsection{Interpretation}
Every binor in $\22^S$ is a state of a
variable subset of $\2S$.

The generator $\gamma_n$ toggles  the $s_n$
into and out of existence.
The uniquadratic property  prevents multiple occupancy of any state.
Like set theory, binor algebra incorporates the Pauli exclusion principle.

The binor product $a\sqcup b = ab$ is the logical XOR operation,
with identity 1 (the empty set) and
vanity
0 (meaningless).
Writing
``$x=0$''  makes it explicit that ``$x$''  is meaningless.
A monomial is a complete description of the variable set,
giving its elements explicitly.
The grade of a monomial is its degree,
the number of factor generators, and the cardinality of that possibility.

The binor sum $+$
is addition (modulo 2)
of polynomials in the
(anti)commuting idempotents $\gamma_k$
 with binary coefficients.
A polynomial, a sum of monomials,
lists possibilities and so gives an incomplete description
unless it is a monomial.

Briefly put, monomials are sharp states,
polynomials are crisp states,
``1'' means ``nothing,'' and
 ``0''  means nothing.

The usual complementof a set is multiplication by the top state,
the product of all the basic generators.

For example
let $a, b, c, \dots$ be distinct generators.  Then the binor product
$abc$ represents the 3-element set with the three unit subsets $a, b, c$.
The sum $a+b+c$ represents the class
of the three possible unit sets $a, b, c$
with equal weights.

If $\gamma$ and $\gamma'$ have grade $g$
then so does $\gamma+\gamma'$.
This sum is only a partial logical operation,  since
$\gamma+\gamma=0$ is the undefined case.
Nevertheless the sum is a group operation
with 0 as its identity,
and the product restricted to monomials is a group operation.

The binor  sum is algebraically similar
to  Boolean addition
modulo 2, but it  is semantically  different:
The binor 0 means nothing while the Boolean 0 means
something, namely  ``nothing."
The quantities combined by the binor sum
are not idempotent but
uniquadratic variables.
Whenever $a= b$\/, the binor sum
$a+b$ is 0, the undefined.
None of the familiar classical logical operations
is undefined in so many cases.

In the Boolean algebra of truth-valued functions $A, B, \dots$ on
a sample space,
with the usual truth-values of 0 (false) and 1 (true),
the following definitions apply, in which all arithmetic is modulo 2,
and the default value in the definitions of POR and PAND  is 0:
\BEA
A\mbox{ OR } B &:=& \mbox{sup}(A, B).\cr
A\mbox{ AND } B &:=& \mbox{inf}(A, B).\cr
A\mbox{ XOR } B &:=& A+B.\cr
A\mbox{ XAND } B &:=& 1+ A + B .\cr
A\mbox{ POR } B &:=& A + B \mbox{ if } A   B  \equiv 0.\cr
A\mbox{ PAND } B &:=& 1+A + B \mbox{ if }  A+B   \equiv 1   .
\EEA
They are listed in dual pairs under complementation,
which replaces every predicate $A$ by its complement $1-A$ ($=1+A$).
POR and PAND are not truth-functional;
that is,
the truth values of $A \mbox{ POR }B$
and $A\mbox{ PAND }B$ at a point are not
functions of the truth values of $A$ and $B$
at that point alone.
Boole and Pierce needed them for probability theory.

The binor algebras are richer languages
than Boolean algebra.
If Boolean algebra represents  mixtures with uniform weight,
 binor algebra
 represents aggregates and powers
as well,
and
the undefined case (0)
as well as the empty case (1).

The  complementation duality of bonor algebra
is top-multiplication
$A\to \top A$.
We extend the above definitions to binor logic
 thus, with 1 now meaning false (absent, the vacuum)
 and $\top$ true (present, the plenum):
\BEA
A\mbox{ OR } B &:=& \mbox{sup}(A, B).\cr
A\mbox{ AND } B &:=& \mbox{inf}(A, B).\cr
 A\mbox{ XOR } B &:=& AB.\cr
A\mbox{ XAND } B &:=& \top AB.\cr
A + B &:=&  A+B.
\EEA
Here the infimum of two binors consists of the monomial
terms present in both,
and the supremum consists of the monomial terms present in either or both.
The sum $A+B$ is self-dual.
To contrast these operations we note that
\BE
1 \;\mbox{XOR}\; 1 = 1\;\mbox{OR}\;1=1,\quad 1+1=0.
\EE
\BE
0\; \mbox{XOR}\; A =  0 \;\mbox{OR}\; A=A;\quad 0+A=A.
\EE

We represent a binor algebra  $\22^{\2S}$ by the formal sum of all its
monomials, which we designate by the same symbol. Then
\BE
\22^{\2S}=\prod_s(1+\iota s)
\EE
 If $\2S=\sum_s s$ has
$N$ terms
then $\22^{\2S}$
has $2^N$ terms.

\subsection{Clifford quantum logic}

We interpret a real or complex
Clifford algebra as a quantum logic in  close
parallel to our interpretation of the binary Clifford algebra.
We use the Clifford algebra as a Hilbert space of state vectors.
Generators $\gamma$ are again sharp specifications
of an individual quantum system \5S.
The Clifford algebra defines
a quadratic form for vectors.
It follows from the Clifford property as usual that
mutually orthogonal vectors  anticommute.
This
 projectively represents the commutative
law of the classical XOR algebra.

Obviously $C$ contains
a representative of every predicate
in the lattice logic of \5S.
These are just the monomial
cliffors that Grassmann called ``real.''

The predicates of $C$ do not represent
filters,
which often stop the system.
They represent phase plates,
which  flip phase rather than stop.
This is a purely
quantum concept.
The
classical irreversible logic arises as a degenerate
form
when one uses
most of the system
as a heat bath and information dump,
making some operations
on the residual system  effectively
irreversible.

By the statistics of the system
we mean the rule for constructing the
algebraic structure of the aggregate system
from that of its constituent.
This is a special case of the process in classical logic
that  the Scottish logician William Hamilton
called quantification in 1842.
This is obviously a more correct term
than ``second quantization'' and  we use it here.

Here
we take
the
the ket space  $H(\5S)$ of the  system \5S
 to be a Clifford algebra:
\BE
H(\5S)={\22}^{V(\6S)}\/.
\EE

\subsection{Hierarchic quantum  logic}

Unlike lattice logic,
Clifford quantum logic iterates neatly to make a hierarchic logic
that is quantum at every level,
and so is at least a candidate for a
higher-order quantum logic.
And unlike Grassmann logic,
which we explored in earlier work \cite{F96},
and  bosonic logic,
which we advocated in one unfortunate paper,
Clifford logic is stable in the Segal sense \cite{SEGAL}
and finite at every finite
order, and
brings in naturally
the indefinite metrics and spinors
that one needs  relativistic gauge theories
like the standard model.
Clifford logic is
the most promising single candidate logic we have found for physics so far.

The Clifford exponential functor
$\Cliff: V \mapsto C= {\22}^V$
 produces an algebra from a quadratic  space.
 We make every Clifford algebra  $C$ a
 quadratic space with the natural
 norm $\|z\|=\Re z^2$ for $z\in C$.
Then Cliff can be iterated
just like the power set functor ${\cal P}: S\to 2^S$.

We designate the vector image in Cliff $C$ of a cliffor $\gamma\in C$
by $\io \gamma$.
$\io$ is isometric:
\BE
\|\io a\| =\Re (\io a)^2 = \Re a^2 =  \|a\|.
\EE

We define the infinite-dimensional hierarchic
Clifford algebra $\Cliff^{\infty}(\iota)$
as the least real Clifford algebra including $\1R$
and closed under $\iota$.
$\Cliff^{\infty}(\iota)$ is the limit (union) of the nested
sequence of
Clifford algebras $C_n$
that starts from the null set for $C_0$
and proceeds by the iteration
\BE
C_{n+1}=\Cliff C_n = \Endo S_{n+1}.
\EE

The first terms in this sequence are
\BEA
C_1&=&\Cliff(0,0)=\1R, \cr
C_2&=&\Cliff(1,0)=\1C,\cr
C_3&=&\Cliff(2,0),\cr
C_4&=& \Cliff(3,1)=\1S,\cr
C_5&=&\Cliff(10,6),\cr
C_6&=&\Cliff(32 832, 32 704),\cr
&\vdots&
\EEA
Here \1S is the 16-dimensional algebra of Dirac spin operators
that Eddington called the sedenion algebra.

The higher algebras $C_7,  C_8, \dots $ have
astronomically large dimensionality $d_n$.
They have the signature
\BE
s_n= \sqrt d_n.
\EE
This is as though the individual generators
had random signatures
$\pm 1$
and the law of large numbers held
exactly for numbers beyond 2, instead of on the average.
We provisionally take
this mathematical hierarchy of quantum logics
for the
physical one
and
use these algebras as basic ingredients in our  constructions.
For example, we use
$C_4$ as the
Dirac algebra of fermion theory.

The hierarchy of sets has a well-known fractal-like  structure.
All the set-making operations that can be applied to the null set
can be applied to any other set.
Thus
growing from every vertex of the tree of sets
is
an image of the entire tree,
and many other sets as well.
$\Cliff^{\infty}(\io)$ has the same fractal-like property.
All the Clifford algebras that occur in it
repeat everywhere in the hierarchy.

\section{Elementary operations}\label{sec:ELOP}

When the state vector space has an algebraic product
expressing statistics we call it the state algebra.
We use the $\iota$ algebra $\Cliff^{\infty}(\iota)$
as state algebra
to describe operations of the cosmic quantum computer \5U.
This provides us with a discrete lexicon of elementary (first-grade)
unipotent operations $o$, $o'$, \dots, ($o^2=1$).
The most general  state vector $|\alpha\ket$ for \5U
is a
polynomial in the elementary operations.

Elementary operations that have been considered (sometimes
implicitly) for the cosmic state algebra include
\begin{itemize}
\item Dirac spins with Maxwell-Boltzmann
statistics (Feynman),
\item Weyl spins with various parastatistics
(Weizs\"acker et al.),
\item Pauli spins with Bose statistics (Penrose),
\item Weyl spins with Bose statistics (Finkelstein),
and
\item iterated Fermi-Dirac statistics (Finkelstein).
\end{itemize}
None of these efforts reached the level of a dynamical theory with
interactions.
We have also considered the 2-valued real Clifford statistics recently.
Two-valued complex representations
of the permutation group were
formulated by Wiman 1898, completely
catalogued by  Schur 1911,
and extensively  studied
in works  such as
Hoffman and Humphreys 1992.
Extended to the orthogonal group,
they begin to be considered for physics in
Wilczek and Zee 1982,
Nayak and Wilczek 1994, and Wilczek 1998.
We do not use them here because they are
not seen in the vacuum.
The single-valued Clifford statistics we use here
reverts to the underlying Clifford logic of
Finkelstein 1982.
We use the Clifford algebra as the state space,
not as
the algebra of observables.
This Clifford statistics is a slight generalization of Fermi-Dirac statistics,
to which it readily specializes.
Particles obeying Clifford statistics we call cliffordons.

\section{Finite  Dirac equation}

We construct an illustrative toy example
of a finite quantum theory of a spin-1/2 particle serving
as a probe to define the space-time structure.
We regularize the theory by replacing compound groups
by nearby simple groups.

We make the following regularization
of
differential geometry within Clifford algebra.
Let  $\mu=1, 2, 3, 4$ be a Lorentz index.
Let $\alpha = 1, \dots, 8$ be an octad index
extending $\mu$.
Let $n=1, \dots , N$ enumerate $N$ octads
of Clifford generators ${\gamma}_{\alpha}(n)$.
For
the space-time differentials we set
\BE
idx^{\mu}\leftarrow {\tau}{\gamma}^{\mu 5}.
\EE
For the space-time coordinates we set
\BE
ix^{\mu}\leftarrow
{\tau} \sum_n
{\gamma}^{\mu 5}(n).
\EE
For the conjugate momenta we set
\BE
ip_{\mu}\leftarrow\frac {n\hbar}{N\tau}\sum_n {\gamma}^{\mu 6}.
\EE
The eigenvalues of the time coordinate  all have the form $n\tau$
for integer $n$ and fall between $\pm N\tau$.
So $\tau$ is a  time quantum or ``chronon.''
There is an analogous energy quantum or ``ergon''
\BE
\epsilon:=\frac{\hbar}{N\tau}.
\EE
For the imaginary unit of finite quantum mechanics,
we infer from the commutation relations that
\BE
i\leftarrow \sum_n
{\gamma}^{56}(n)/N=:\eta
\EE
This $\eta$ is an operator in the algebra
and hence a quantum variable.
In
quaternionic quantum mechanics
(Finkelstein Jauch and Speiser 1959)
the variable $\eta=\eta(x)$ is a natural (St\"uckelberg-) Higgs field;
indeed, this is still the only
Higgs field since gravity that was not introduced ad hoc.
Clifford algebra is a generalized
quaternion algebra,
and it is natural optimism to expect the variable $\eta\in \22^{8N\1R}$
to serve as Higgs field once again.
The $i$ given above, however,  has a rich spectrum of values in $[0, 1]$
for its absolute value.
To account for the observed near-constancy of $i$
(as the transformation from anti-symmetric symmetry generators to
symmetric observables)
we must appeal to a long range order
among all $N$ of the ${\gamma}^{56}(n)$
in the vacuum ground state;
much as if the usual $i$ is the effective vacuum value of a Higgs field.

Using these correspondences
it is easy to reconstruct  the Dirac equation for a
neutral particle as a correspondence limit of
a finite Clifford algebraic quantum theory.
We suppose that the rest mass $m$ is
the correspondence limit of a conjugate variable to a proper time variable
$\tau$
\BE
m= i\hbar {d \over d\tau}\/.
\EE
We  assign the
proper-time coordinate $\tau$ to $\5T$, not $\5S$.
From the viewpoint of the CE, they are all \5U variables.

The Dirac equation that we have to regularize is then the operator equation
of motion for any one-particle  operator $X$, not explicitly depending
on ${\tau}$, based on a proper-time generator $-iM$:
\BE
{d\over d{{\tau}}} X = -i[M, X]
\EE
with mass operator
\BE
M= {\gamma}^{\mu 5}\pd_{\mu}.
\EE

The flexed algebra is a large Clifford algebra $C=\22^V$ over a
quadratic space $V=8N\1R$.
$2^V$ is a Clifford product of $N$ octadic Clifford algebras
$\22^{\28}$.
This is isomorphic by a Jordan-Wigner transformation to a tensor product
of
$N$ octadic Clifford algebras:
\BE
\22^{N\28}\cong \prod_{n=1}^N \22^{\28} \cong \OX _{n=1}^N \22^{\28}
\EE
which represents a
Maxwell-Boltzmann aggregate of octads.
This is how we  account for the fact that
space-time points have
Maxwell-Boltzmann statistics in
the standard physics.

This calculation ultimately
has to be made
self-consistent, as
whenever one postulates
a spontaneous symmetry breaking.
We must show that the dynamics can
lead to a ground mode with the spontaneous symmetry breaking
by $\eta$ that we have assumed.

This project further requires flexing the gauge group of present
differential-geometric physics
to arrive at a Clifford algebraic theory with interaction
This is one of several problems now being studied.

Since this meeting a finite quantum  linear harmonic oscillator has been worked out
\cite{MOHSEN}/

\section{Acknowledgments}
We gratefully acknowledge
 L. Susskind  and F. Wilczek
for  helpful and stimulating discussions,
and the M. \& H. Ferst Foundation and the Elsag Corporation for important
support.

\tableofcontents

\end{document}